\documentclass[aps,english,showpacs,10pt]{revtex4}

\usepackage[latin1]{inputenc}
\usepackage{float}
\usepackage{latexsym}
\usepackage{graphicx}
\usepackage{amssymb}
\usepackage{amsmath}
\usepackage{amsfonts}
\usepackage{hyperref}
\usepackage{color}
\usepackage{cool}

\newcommand\be{\begin{equation}}
\newcommand\ba{\begin{eqnarray}}
\newcommand\ee{\end{equation}}
\newcommand\ea{\end{eqnarray}}

\begin{document}

\title{Holographic Curvature Perturbations in a Cosmology with a Space-Like Singularity}
\author{Elisa G. M. Ferreira}
\email{elisa.ferreira@mail.mcgill.ca}
\affiliation{Department of Physics, McGill University, Montr\'eal, QC, H3A 2T8, Canada}
\author{Robert Brandenberger}
\email{rhb@physics.mcgill.ca}
\affiliation{Department of Physics, McGill University, Montr\'eal, QC, H3A 2T8, Canada, 
and Institute for Theoretical Studies, ETH Z\''urich, CH-8092 Z\''urich, Switzerland}
\date{}

\begin{abstract}

We study the evolution of cosmological perturbations in an anti-de-Sitter (AdS) bulk
through a cosmological singularity by mapping the dynamics onto the boundary
conformal fields theory by means of the AdS/CFT correspondence. We consider
a deformed AdS space-time obtained by considering a time-dependent dilaton
which induces a curvature singularity in the bulk at a time which we call $t = 0$,
and which asymptotically approaches AdS both for large positive and negative
times. The boundary field theory becomes free when the bulk curvature goes to infinity. 
Hence, the evolution of the fluctuations is under better controle on the boundary than 
in the bulk. To avoid unbounded particle production across the bounce it is necessary to 
smooth out the curvature singularity at very high curvatures. We show how
the bulk cosmological perturbations can be mapped onto boundary gauge field
fluctuations. We evolve the latter and compare the spectrum of fluctuations on
the infrared scales relevant for cosmological observations before and after the
bounce point. We find that the index of the power spectrum of fluctuations
is the same before and after the bounce.
 
\end{abstract}

\maketitle

\section{Introduction}

In spite of its many successes, cosmology faces a number of outstanding theoretical and 
fundamental challenges. One of the most fundamental questions that remain to be answered 
in cosmology is the singularity problem. Singularities appear in many cosmological models 
and are unavoidable in some contexts like Einstein gravity with matter fields that do not 
violate the null energy condition (NEC). Both Standard Big Bang cosmology and the inflationary
universe scenario \cite{Guth} realized in the context of scalar field matter coupled to classical General
Relativity are examples where an initial Big Bang singularity is present (see the
classic paper \cite{Hawking} for the proof that an initial singularity appears in
Standard Big Bang cosmology and \cite{Borde} for an extension to inflationary
cosmology). Bouncing cosmologies, alternative scenarios for the evolution of the universe 
where a contraction period precedes the expansion of the universe, also have singularities
at the bounce point, at least if they are realized within the realm of Einstein or dilaton gravity
coupled to matter obeying the NEC. One can avoid Big Bang/Big Crunch singularities 
by postulating matter which violates the NEC (see e.g. \cite{quintom, ghost, Galileon} for some
specific models, and \cite{RHBmbRev} for a review), or by going beyond, Einstein gravity
(e.g. by choosing gravitational Lagrangians with specifically chosen higher derivative
terms \cite{BM}, by considering the Horava-Lifshitz gravitational action \cite{HLbounce}, or
by assuming certain nonlocal gravitational Lagrangians \cite{Biswas}). However, there
are doubts as to whether these constructions can be embedded in a consistent quantum
theory of gravity \cite{Adams}. A consistent understanding of singularity resolution can 
presumably only be studied in such an ultraviolet complete theory, superstring theory
being the prime example.

In this context the AdS/CFT correspondence could come to use. This correspondence 
\cite{Maldacena} is a proposal for a non-perturbative treatment of string theory 
and states that the dynamics of a bulk Anti-de Sitter (AdS) space-time that includes 
gravity is encoded in the boundary of this space-time, where a conformal field theory (CFT) 
with no gravity lives. This conjecture has been used in many different fields of 
physics, from black hole physics to condensed matter, with great success (for a review 
e.g. \cite{AdS-CFTrev}). It has already been proposed in the literature that the 
AdS/CFT correspondence could be used to resolve cosmological singularities
\cite{HH,CHT,CHTproblem,CHT2}, specially in the context of singular bouncing models
such as the Pre-Big-Bang \cite{PBB} and Ekpyrotic \cite{Ekp} scenarios. 

Bouncing cosmologies have recently been studied extensively as possible alternatives
to cosmological inflation for producing the fluctuations which we are currently mapping
out with observations. If the equation of state of matter in the contracting phase has
$w > - 1/3$, where $w$ is the ratio of pressure to energy density, then scales exit
the Hubble radius during contraction. Hence, it is possible to have a causal generation
mechanism for fluctuations in the same way as in inflationary cosmology, where
scales exit the Hubble radius in an expanding phase if the equation of state of
matter obeys $w < - 1/3$. As was pointed out in \cite{FB, Wands}, if the equation
of state of matter during the time interval when scales which are measured now
in cosmological observations exit the Hubble radius has the equation of state
$w = 0$ (i.e. a matter-dominated equation of state), then initial vacuum perturbations
originating on sub-Hubble scales acquire a scale-invariant spectrum, the kind of
spectrum which fits observations well \footnote{A slight red tilt of the spectrum emerges
if the effects of a dark energy component are included \cite{Yifu}.}. A scale-invariant
spectrum of fluctuations can also be obtained in the Pre-Big-Bang \cite{PBBsi} and in the
Ekpyrotic \cite{NewEkp} scenarios, making use of entropy modes. The major problem
in these analyses is that the fluctuations have to be matched from the contracting phase
to the expanding phase across a singularity (for singular bouncing cosmologies) or
in the region of high curvature (in nonsingular models in which new physics provides
a nonsingular bounce) where the physics is not under controle. This is the second
place where the AdS/CFT correspondence could become useful: the boundary
theory becomes weakly coupled precisely where the bulk theory becomes strongly
coupled, and hence we can expect that the evolution of the fluctuations on the
boundary will be better behaved.


We here consider a time-dependent deformation of AdS \cite{DT1, DT2, DT3, DT4}
(see also \cite{CH1, CH2}) which yields a contracting phase with increasing
curvature leading to a bulk singularity at a time which we call $t = 0$. The evolution
for $t > 0$ is the mirror inverse of what happens for $t < 0$. This means that
the bulk is expanding with decreasing curvature.
The challenge for our work hence is to explore if the AdS/CFT correspondence 
can be used to determine the cosmological perturbations in the expanding phase  
starting with some initial cosmological perturbations in the contracting phase. In
the case of a singular bouncing bulk cosmology this question cannot be answered 
from the point of view of the bulk evolution of those perturbations, and in a
non-singular bouncing setup the evolution in the bulk cannot be reliably
computed in a perturbative approach. For example, there are ambiguities
if one wants to apply the matching condition approach \cite{HV, DM} to connect early
time to late time fluctuations (see e.g. \cite{DV}). The goal of our work is to avoid
these difficulties in the bulk evolution in the strongly coupled region by mapping
the dynamics onto the boundary theory which is weakly coupled near $t = 0$.
The AdS/CFT correspondence presents an unique opportunity to understand 
the effects of a bulk singularity on cosmological observables. Specifically,
we are interested in computing the amplitude and slope of the spectrum of
cosmological perturbations after the bounce given the spectrum before the
bounce.
 
In \cite{Brandenberger:2016egn} the authors studied the evolution of matter scalar 
field perturbations using the AdS/CFT correspondence in a deformed AdS$_5$ spacetime, 
where a spacelike singularity is present. This background spacetime is a time 
dependent background studied before in \cite{DT1,DT2,DT3,DT4} where the dilaton 
bulk field has a time dependence which as $t \rightarrow 0$ produces a curvature 
singularity. The bulk theory is weakly coupled for $|t| > t_b$ and
strongly coupled for smaller values of $|t|$. In the context of this background the authors 
studied dilaton perturbations on a hypersurface perpendicular to the AdS radial coordinate, 
starting with a scale invariant spectrum on super-Hubble scales at early times $t < - t_b$. 
When bulk gravity becomes strongly coupled at $t = - t_b$, the perturbations 
were mapped to the boundary theory, a $\mathcal{N} = 4$ Super Yang-Mills (SYM) 
theory, and the fluctuations of the corresponding boundary fields were then
evolved from $t = -t_b$ to $t = t_b$. This SYM model has a time-dependent coupling 
constant that goes to zero at the same time as the singularity occurs in the bulk. However,
in spite of the fact that the boundary theory becomes free at $t = 0$, it was found that
infinite particle production occurs between $t = - t_b$ and $t = 0$. Thus, it was
necessary to introduce a cutoff: the coupling constant was kept finite but small in a 
short period of time $|t| < \xi$ around the singularity, where $\xi \ll t_b$. This 
made it possible to evolve the fluctuations unambiguously past the time $t = 0$ where
the bulk singularity occurs until the late time $t = t_b$ after the singularity
after which the bulk theory becomes weakly coupled again. After that, for the infrared 
modes that are of cosmological interest (and whose wavelength is much larger
than the Hubble radius already at the time $t = - t_b$), the bulk scalar field perturbations were 
reconstructed. It was found that the late time scalar field perturbations have a 
scale invariant spectrum, showing that the spectral index does not change while 
passing through the region of the highly curved (and maybe even singular) bulk. 
On the other hand, the amplitude of the scalar field perturbation spectrum is
amplified - a consequence of the squeezing of the perturbation modes on
super-Hubble scales in the contracting phase.
 
The evolution of scalar matter perturbations is interesting since it offers us a
good guide as to the evolution of gravitational waves \footnote{The squeezing
of the amplitude of gravitational waves on super-Hubble scales is governed
by the same equation as the squeezing of matter scalar field fluctuations,
whereas the scalar metric fluctuations are in general squeezed by a different
factor - see e.g. \cite{RHBfluctrev} for a short review, and \cite{MFB}
for a more comprehensive survey of the theory of cosmological perturbations.}. 
However, of more interest in cosmology is the spectrum of the scalar metric perturbations, 
since those lead to the adiabatic density perturbations responsible for structure formation 
in the universe. The goal of the present paper is to extend the analysis
of \cite{Brandenberger:2016egn} to the case of cosmological perturbations.

Scalar cosmological perturbations are more complicated to analyze than
matter scalar field fluctuations. They are made up of a combination of metric
and matter inhomogeneities which take different forms in different coordinate
systems. In the case of purely adiabatic perturbations \footnote{For a single matter
field the perturbations on super-Hubble scales are automatically adiabatic. In the
case of multiple matter fields the adiabaticity condition means that the relative
energy density fluctuations in each matter field are the same.} the information
about the inhomogeneities is most conveniently encoded in the quantity 
$\mathcal{R}$, the curvature perturbation in comoving gauge (the gauge in
which the matter field fluctuation vanishes \cite{Bardeen}), a quantity that remains 
constant in time outside the Hubble radius \cite{BST, BK, Lyth, AB, LV}. 

According to the AdS/CFT dictionary, the metric perturbation $\delta g_{\mu \nu}$
has as its dual operator in the CFT  the expectation value of the boundary 
energy momentum tensor. In order to  reconstruct the curvature perturbations 
(which are a combination of the metric and the matter fluctuations)
in the future of the space-time singularity, one needs 
to know the full evolution of the boundary operators corresponding to 
both the bulk matter scalar field and the metric perturbations. 

We argue in this paper that there exists a gauge choice that can simplify this problem. 
We generalize the spatially flat gauge to the $5$-dimensional case. This gauge allows 
us to describe the curvature perturbations as a function only of the perturbations of the 
scalar field, that represents the matter in our space-time. We do this by using the 
gauge freedom to gauge away the metric perturbations degrees of freedom which
leaves us with only the scalar field perturbations. Because of this choice of gauge we
only need to know how the scalar field perturbations behave at late times. 
This allows us to perform the same analysis as in \cite{Brandenberger:2016egn}, and to 
evolve the perturbations of a scalar field using the boundary theory in a singular deformed 
AdS$_5$ space-time. With this gauge choice, the analysis of scalar field
fluctuations is all we need to be able to reconstruct the curvature perturbations 
in the future of the space-time singularity.

This paper is organized as follows. Section II contains a summary of the dynamics 
of the deformed AdS bulk space-time containing a spacelike singularity. In Section III 
we discuss the cosmological perturbations in this deformed AdS$_5$ space-time and
present the generalized spatially flat gauge. Section IV shows the main result,
namely how to obtain the curvature perturbation at late times evolving from an 
initial bulk perturbation using the AdS/CFT correspondence. We find that the 
spectral index is not changed when comparing the spectrum at late and early
times, but that there is an increase in amplitude resulting from the squeezing
of the Fourier mode wave functions.

\section{Bulk Dynamics}

We are interested in studying cosmological backgrounds in the context of the 
AdS/CFT correspondence.  Some time-dependent backgrounds in string theory 
were studied in \cite{DT1,DT2,DT3,DT4} where the bulk solution can be thought as a 
time-dependent deformation of $AdS_5 \times S^5$ with a corresponding 
$\mathcal{N}=4$ supersymetric Yang-Mills (SYM) theory with a time dependent 
gauge coupling constant as a dual theory.

This background bulk solution can be described by the line element
\begin{eqnarray}
\label{eq:gBG}
  ds^2 &=& \frac{\ell^2}{z^2}\left( dz^2 + \tilde{g}_{ab} dx^a dx^b \right) \nonumber \\
  &=& \frac{\ell^2}{z^2}\left[ dz^2 + \frac{2 |\tau|}{3 \ell} \left(-d\tau^2 + \delta_{ij} dx^i dx^j\right) \right] \,, \quad 
  \quad t \gtrless 0, \; z \in (0,+\infty)\,,
\end{eqnarray}
plus a $d\Omega ^2 _5$ term representing the $S^5$ factor.  In the second line we 
are choosing a special Kasner type solution. Note that $\tau$ denotes
conformal time.The dilaton profile is given by
\begin{eqnarray}
\label{eq:phiBG}
  \phi(x) = \phi(t) = \sqrt{3} \ln\left(\frac{|t|}{\ell}\right) \,, \quad
  \Rightarrow \quad
  e^{\phi(x)} = \left(\frac{|t|}{\ell}\right)^{\sqrt{3}} \,,
\end{eqnarray}
with $\ell$ being the AdS scale. Throughout this paper we use conventions that 
Greek letters $\mu,\nu,\dots$ run over all of the five space-time indices
$0,\dots 4$ with $z$ the AdS radial dimension; latin indices from the beginning 
of the alphabet $a,b,\dots$ run over the indices $0,\dots,3$ corresponding
to the four-dimensional space-time perpendicular to the AdS radial direction,  
and Latin letters $i,j,\dots$ run over the spatial indices $1,2,3$.

This solution can be embedded in a solution of a $10$-dimensional type IIB 
supergravity theory provided that the metric and the dilaton satisfy the equations 
of motion
\begin{equation}
\partial_{\mu} \left( \sqrt{-\tilde{g}} \tilde{g}^{\mu \nu} \partial_\nu  \phi (x) \right) = 0, 
  \quad 
  R_{ab}[\tilde{g}] = \frac{1}{2} \nabla_a \Phi \nabla_b \phi(x) \,.
\end{equation}
The bosonic sector of this embedding includes a RR $5$-form flux which 
supports the $S^5$ tensor factor of the space-time. The  $S^5$ factor will not 
play a role in the following and we will thus not track it further. 

\begin{figure*}[t]
\begin{center}
\includegraphics[scale=0.3]{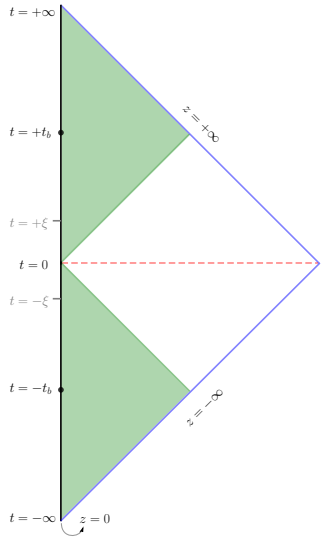}
\caption{Conformal diagram of the background space-time.
 The spacelike singularities is shown in red. The vertical axis is time $t$, the horizontal direction
represents the coordinate $z$ (with the boundary on the left
side). If there were no deformation of $AdS$, the region drawn
would correspond to the Poincare patch of $AdS$. The green regions
may be covered by Fefferman-Graham \cite{FG} charts with Minkowski boundary metrics. }
\label{background}
\end{center}
\end{figure*}

The element (\ref{eq:gBG}) is easily recognized to be a deformation of the line element 
of pure AdS in Poicar\'e coordinates, where the AdS coordinate runs from $z=0$ at the 
boundary to $z=\infty$ at the Poincar\'e horizon. In pure AdS the induced metric on 
constant-$z$ hypersurfaces is the Minkowski metric. In our solution the induced 
metric is instead composed of two copies of a Friedmann-Robertson-Walker (FRW) 
metric, as seen in (\ref{eq:gBG}), one for $t<0$ describing a collapsing geometry, 
and another for $t > 0$ describing an expanding geometry. 
The solution contains a spacelike singularity, "Big Crunch" singularity, at $t=0$. 
It is also singular as $z \to \infty$ at any fixed $t \neq 0$. This can be seen in 
Figure ~\ref{background}. Due to this latter fact, the spacetime cannot be 
Cauchy-extended beyond the Poincar\'e horizons which bound the coordinate chart. 
The string coupling is given by $g_s=e^{\phi(x)}$ and goes to zero at the singularity.
If the singularity can be resolved by mapping the dynamics to the boundary, we will 
have a stringy realization of a boucing scenario.

We are assuming the the bulk universe initially begins in an AdS vacuum 
at some early moment $t_i$ and the background is given by a weakly coupled
supergravity theory. At the moment $- t_b$, the bulk gravity becomes strongly
coupled but at the same time the boundary gauge theory become weakly coupled. 
Hence, after the time $- t_b$ the evolution on the boundary becomes tractable
in perturbation theory. On the future side of the
bulk singularity, the boundary theory remains tractable perturbatively until
the time  $t_b$ when the the bulk theory becomes weakly coupled
again at the cost of the boundary theory becoming strongly coupled. At
that time we can reconstruct the bulk information (at least in the vicinity of the 
boundary) from boundary data (see e.g. \cite{Hamilton1, Hamilton2, Kabat, Ian}).

We will take our space-time to be the hypersurface of some constant AdS
radial coordinate $z$. We will be interested in considering linear fluctuations
of matter and scalar metric degrees of freedom on this surface at some
initial time $t \ll - t_b$, and computing the corresponding fluctuations on
the same constant $z$ surface in the future of the singularity, once the
bulk theory once again becomes weakly coupled, i.e. at $t = t_b$.

To resolve the singularity in the background (\ref{eq:gBG}, \ref{eq:phiBG}) 
and study the evolution of perturbations to the future of $t = - t_b$ we will 
map the problem onto the boundary using the AdS/CFT 
dictionary. We can see that the boundary of this $5$ dimensional solution is 
conformally flat and has a second order pole as $z\rightarrow 0$. So, in order 
to do holography we must specify a conformal frame, \textit{i.e.}~we must 
provide a defining function $\Omega(x)$ which behaves like $O(z^2)$ as 
$z \to 0$, and which in turn selects the induced boundary metric $h_{ab}(x)$ via
\begin{eqnarray}
\label{eq:h}
  ds^2_{bndry} := h_{ab}(x) dx^a dx^b 
  = \lim_{z \to 0} \Omega^2(x) \gamma_{ab}(x) dx^a dx^b \, ,
\end{eqnarray}
where $\gamma_{ij}$ is the metric of a maximally symmetric three-dimensional
hypersurface (the metric of Euclidean three space, of the three sphere or
the three-dimensional hypersphere).
The above is an asymptotic solution in Fefferman-Graham form \cite{FG} that 
represents the conformal structure. There are two natural choices for 
$\Omega(x)$. First, if we select
\begin{equation}
  \Omega_{\rm FRW}(x) = \frac{z}{\ell},
\end{equation} 
then the boundary limit is particularly simple and the conformal boundary has 
metric $h_{ab}(x) = \gamma_{ab}(x)$. We refer to this as the FRW frame. 
A second choice is 
\begin{equation}
  \Omega_{\rm M}(x) = \left(\frac{3 z^2}{2 \ell |t|}\right)^{1/2} \, .
\end{equation} 
With this choice the boundary metric is flat. We refer to this as the Minkowski frame. 
It is important to realize that $\Omega_M(x)$ is singular as $t \to 0$, and as a 
result the conformal transformation implied by $\Omega_M(x)$ is singular. 
One of our basic assumptions is that this conformal transformation is nevertheless 
a symmetry of the CFT.

Let us now turn to the CFT description of our solution. In this time-dependent background, 
the dual  boundary theory is a $\mathcal{N}=4$ SYM theory with a source. Following the 
usual dictionary \cite{Skenderis:2000in},
the AdS-Neumann part of $\phi(x)$ sets the value of the Yang-Mills coupling via 
\begin{equation}
  g^2_{YM}(x) = \lim_{z\to 0 } e^{\phi(x)} \, .
\end{equation}
In the Minkowski frame, this theory lives in flat space.
Note that when the non-normalizable part of $\phi(x)$ varies with time, as in our example, 
the SYM theory in the boundary is sourced by a coupling that is time-dependent. So, the 
time dependent dilaton in the bulk corresponds to a time-dependent Yang-Mills coupling 
in the boundary. This coupling goes to zero as $t\rightarrow 0$ and the CFT becomes free.

\section{Cosmological Perturbations in the Deformed AdS$_5$}

We want to compute the cosmological perturbations from the space-time described 
above. In \cite{Brandenberger:2016egn}, we perturbed only the scalar field, namely, 
the dilaton, and treated the perturbations of the dilaton. However, to fully describe the 
cosmological perturbations, we need to include the perturbations of the metric. For 
this, we need to perturb this deformed AdS$_5$ metric (see e.g. \cite{5dflucts} for
general discussions of cosmological fluctuations in brane world like five dimensional
space-times). 

Our starting point is the perturbed five-dimensional space-time metric
\begin{equation}
g_{MN} \, = \, g_{MN}^{(0)} \, + \, \delta g_{MN}\,,
\end{equation}
where the first term on the right hand side denotes the background metric
which depends only on $t$ and $z$, and the second the linear fluctuations
(which depend on all five space-time coordinates). 

We can make a field redefinition in order to write the 
background metric in the following way:
\begin{equation}
ds^2 \, = \, d\bar{z}^2 + \tilde{g}_{ab} d\bar{x}^a d\bar{x}^b\,,
\end{equation} 
where $\bar{z}=(l/z)z$ and $\bar{x}^{\mu}=(l/z)x^\mu$. In some papers in the 
literature this is called a Gaussian normal coordinate system, and 
it is equivalent to restrict the coefficient in front of the $z$-part of the metric to 
be unity. In the following we will drop the tilde signs on the
coordinates.

After including linear fluctuations, the metric can be written,  performing the usual
scalar-vector-tensor decomposition with respect to spatial rotations on the
constant $z$ spatial hypersurface, as (see e.g. \cite{MFB, RHBfluctrev})
\begin{equation}
ds^2 \, = \, - dz^2 (1 + C_0) + C_adx^a + a^2(\tau) \left[ \left( 1+2\Phi \right) d\bar{\tau}^2 - 
2B_{i} d\bar{x}^i d\bar{\tau}- \left( \delta_{ij} + h_{ij}  \right) \right]dx^i dx^j \, , 
\end{equation}
where $\tau$ is the conformal time given by $d\tau = dt / a(t)$,
and $\Phi, B_i$ and $h_{ij}$ are functions of all space-time variables. 
The linear quantities $C_0$ and $C_a$ are new metric fluctuations
associated with the presence of the radial AdS direction. We can further 
decompose the $3-$vector $B_i$ into a scalar and a divergenceless part and the 
rank-$2$ symmetric tensor $h_{ij}$ into scalar, a vector and tensor parts:
\ba
B_i \, &=& \, \partial _i B + \hat{B}_i \\ 
h_{ij} \, &=& \, 2\Psi \delta _{ij} + 2 \left( \partial_i \partial_j  - \frac{1}{3} \delta_{ij} \nabla^2\right) E 
+ \left( \partial_i \hat{E}_j + \partial _j \hat{E}_i \right) + 2 \hat{E}_{ij}\,, \nonumber
\ea
where this decomposition is irreducible since the hatted quantities are divergenceless, 
$\partial^i \hat{E}_i$ and $\partial^i \hat{E}_{ij}=0$, and the tensor part is traceless, 
$\hat{E}^i _i=0$. Note that the tensor $\hat{E}_{ij}$ corresponds to gravitational
waves, $\hat{B}_i$ and  $\hat{E}_j$ to vector perturbations, and the remaining
functions $\Phi, \Psi, B$ and $E$ to the scalar metric perturbations. This perturbed 
metric and the variables are analogous to the 
fluctuations in a usual $4$-dimensional cosmology when restricted to constant 
$z$ slices. However, one needs to remember that the quantities calculated also 
depend of the coordinate $z$. In the following we will neglect vector perturbations
and gravitational waves. 

Together with the metric perturbations, we need to perturb the energy-momentum 
tensor of the $5$-dimensional bulk. The matter content in our case is the dilaton field. 
This can be perturbed as follows, as in \cite{Brandenberger:2016egn}
\begin{equation}
\bar{\phi} \left( z,\mathbf{x},t \right) \, = \, \phi \left( z,\mathbf{x},t \right) 
+ \delta \phi \left( z,\mathbf{x},t \right)\,,
\end{equation}
where $\phi$ represents the background dilaton field and $\delta \phi$ its linear perturbation. 

General relativity allows for a freedom in the choice of the coordinate system.  At
the linearized level the space of coordinate transformation is five-dimensional,
allowing us to impose five gauge conditions in order to remove residual
gauge degrees of freedom. As is done in the four space-time dimensional theory
of cosmological perturbations we use two of these gauge freedoms to simplify
the scalar sector of the metric. One choice is {\it longitudinal} gauge in which
one sets $B = E = 0$. Two gauge degrees of freedom are vector from the point
of view of the constant $z$ hypersurfaces and can be used to reduce the
number of vector modes, and the remaining gauge degree of freedom involves
the $z$ direction and could be used to set $C_0 = 0$. Making these choices,
the scalar cosmological fluctuations involve the variables $\Phi, \Psi$ and $\delta \phi$
(plus the variables $C_a$ which will not be important for us). An alternative
choice is to pick the \textit{spatially flat gauge} (\textit{uniform curvature gauge})
in which the curvature on the constant time (and $z$) hypersurfaces is constant
in space. In the absence of anisotropic stress $\Phi$ and $\Psi$ coincide,
and the Einstein constraint equation related the other two variables \footnote{An
easy way to see this is by noting that a matter perturbation $\delta \phi$
inevitably leads to a metric fluctuation of scalar type.}. Hence, on a fixed
$t$ and $z$ hypersurface, the information about scalar cosmological perturbations
is encoded in terms of a single function.

Our goal will be to compute the evolution of the $3 + 1$ dimensional curvature fluctuation variable
$\mathcal{R}$,  which in the absence of entropy fluctuations is conserved on 
super-Hubble scales and thus encodes the relevant information about the
scalar cosmological perturbations. It is hence the useful variable to track on super-Hubble
scales, the scales we are interested in in this work (and also the ones which
are of interest in inflationary cosmology).

We choose to work  in uniform curvature gauge. In this gauge, the variable
$\mathcal{R}$ is on super-Hubble scales given by
\begin{equation}
\mathcal{R} \, = \, -\frac{\mathcal{H}}{\phi'} \delta \phi\,.
\label{curvature}
\end{equation}
in terms of the scalar field fluctuation $\delta \phi$. The
coefficient relating $\mathcal{R}$ and $\delta \phi$ is given by
the comoving Hubble constant and by the background scalar field. Note that a
prime indicated the derivative with respect to conformal time.

The variance for $\mathcal{R}$ in this gauge is given by the variance of $\delta \phi$.
For each Fourier mode we have
\begin{equation}
\langle |\mathcal{R}_k|^2 \rangle \, = \, 
\left( \frac{\mathcal{H}}{\phi '} \right) ^2 \langle |\delta \phi_k|^2 \rangle \, .
\end{equation}

\section{Holographic Curvature Perturbation at Late Times}

The goal of the section is to calculate the conserved curvature perturbation in our 
deformed spacetime at late times. With the general spatially flat gauge developed 
above, we are able to write the curvature perturbation in terms of the perturbations 
of the scalar field and its background value. This is important in our setup, since 
it avoids one having to understand how the metric perturbations evolve 
holographically in this singular spacetime.

In our previous work \cite{Brandenberger:2016egn}, we showed a prescription for 
obtaining the bulk perturbation of a scalar field $\delta \phi$, the dilaton in our case, at 
late times $t > t_b$ in the weakly coupled region of the expanding
period, given initial conditions for the perturbations of the scalar field in the bulk in 
the weakly coupled contracting phase. We were interested to understand how the 
presence of the singularity affects the initial scalar field  perturbations. In particular, 
we investigated if the power spectrum given in the bulk at past times is changed  
after the singularity. We showed that the final spectrum of $\delta \phi_k$ remains 
scale invariant, given it was scale invariant in the past.

We now show how to use the results of our previous work to
compute the quantity that is of interest in cosmology, the curvature perturbation. 
As we saw in the previous section, then when working in the \textit{uniform spatial curvature gauge}
we only need the power spectrum of the scalar field to obtain the power spectrum
$P_{\mathcal{R}}$ of the curvature perturbations
\begin{align}
P_{\mathcal{R}} \, = \, \left( \frac{\mathcal{H}}{\phi '} \right) ^2 P_{\delta \phi} = \frac{1}{2\pi ^2} k^3 \left( \frac{\mathcal{H}}{\phi '} \right) ^2 |\delta \phi _k \left( z,t \right)| ^2 \,.
\label{power_spectrum}
\end{align}
Thus, if we have the solution of the bulk scalar field in the future of the singularity, 
$\delta \phi_k (+t_b)$, we are able to obtain the power spectrum of the curvature perturbation 
from the above relation. 

As it was shown in \cite{Brandenberger:2016egn},  the bulk
scalar field fluctuations in the future of the singulariy can be locally
reconstructed from the boundary data via \cite{Hamilton1,Hamilton2}
\begin{equation}
\delta \phi \left( t, x, z \right) \, = \, \int dt' d^3y K \left( t',y | t, x, z \right) \mathcal{O} \left( t',y \right)\,,
\label{reconstruction0}
\end{equation}
where the kernel $K \left( t',y | t,x,z \right)$ is the smearing function. The
important property of this smearing function is that its support is confined to the ``AdS causal
wedge'', {\it i.e.} to points with $|\delta t| \equiv t^{\prime} - t < z$ (see Fig. 2). The
exact form of the smearing function will not be important for our analysis.This construction 
is similar to a boundary value problem (see also \cite{Ian, Alberto}). 
This means that $\delta \phi  \left( t, x, z \right)$ 
corresponds to a local operator $\mathcal{O} \left( y, t' \right)$ in the CFT, with a map 
defined by the smearing function. 

\begin{figure*}[t]
\begin{center}
\includegraphics[scale=0.4]{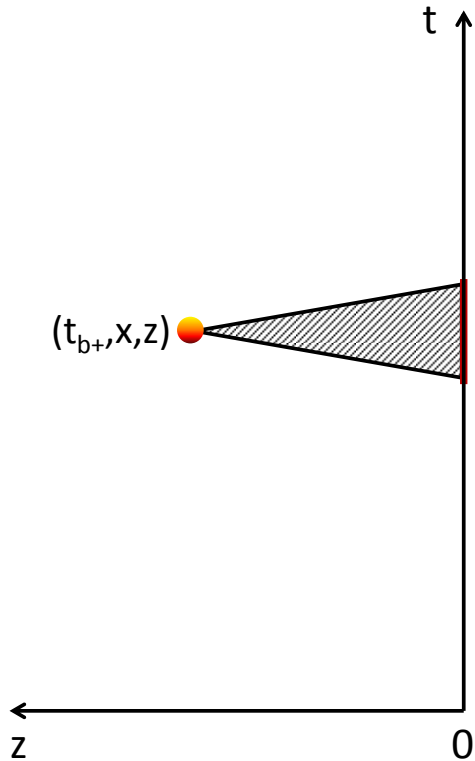}
\caption{Sketch of the reconstruction process for bulk operators
in the future of the singularity. The vertical axis is time,
the horizontal axis indicates the radial coordinate $z$.
The solid line is the boundary.The bulk field at a point with 
coordinates $(t_b^+, x, z)$ is given by integrating the
boundary data against the boundary-to-bulk propagator. The
integration involves data in the shaded region only. The solid curves
connecting the bulk point to the boundary are null geodesics.}
\end{center}
\end{figure*}

We can make a translation in the time and space coordinates: $t'=t+s$ and $y=x+y'$. 
The smearing function is invariant under translations of the $x$ coordinates. In pure AdS
it would also be invariant under time translations. In the case of our deformed AdS,
the kernel has an explicit time-dependence. The important point is that the kernel
has support within the AdS causal wedge. Hence, as long as we consider values
of $z$ not too far from the boundary, the region of support of the kernel for 
the time $t = t_b$ is far away from the space-time singularity, and the
kernel is hence well defined. The fact that the kernel is independent
of the three dimensional spatial coordinates $x$ on the fixed $z$ surface
implies that the kernel does not effect the shape of the power spectrum.

Our interest is to be able to find the spectrum of the perturbations in the future. For that, 
we need to work with the Fourier transform of (\ref{reconstruction0}). This is given by:
\ba
\delta \phi _k \left( z,t \right) \, &=& 
\int ds d^3 y' e^{-iky'}  K \left( t + s, y' | z, t  \right) \mathcal{O} \left( k, t+s \right) \,,  \nonumber \\
&\simeq& \int ds \underbrace{\int d^3 y' K \left( s + t, y' | z, t \right)}_{M(t, s,z)} \mathcal{O} \left( k, t +s \right)
\label{reconstruction}
\ea
where in the last line we considered only the IR limit, the one of interest for cosmological 
perturbations. We can see from this equation that  $\mathcal{O} \left( k, t +s \right)$ has the 
same $k$-dependence of $\delta \phi _k$, with the amplitude smeared and calculated 
at a translated time. 

We do not need to know the exact form of the kernel for our deformed AdS$_5$. All
we need to assume is the existence of such a smoothing function with a causal structure
similar to the one for pure AdS (obtained in \cite{Hamilton1,Hamilton2}). The differences 
would appear in the time-mode solution, since here we have FRW spacetime on the 
boundary instead of Minkowski, leading to a different normalization for $K(s, y'| z)$. 
Also, this function has support on the causal wedge of $AdS$, and selects only data 
on the boundary that is space-like separated with $\mathbf{ky'} << 1$, given by the 
values of $\mathbf{y'}$ where the smearing function does not vanish. 
It is very important in our case that the only data necessary for the reconstruction of the
bulk field is local, since the presence of the singularity makes part of the data in the past 
inaccessible in the future of the deformed AdS Poincare chart. We can see that from the 
green regions of Figure \ref{background}.

We can then write the power spectrum of the uniform curvature perturbation with 
respect to the boundary data. From equation (\ref{power_spectrum}), and knowing 
that for our bulk $\mathcal{H} = a^{\prime}/a$, where 
\be
a(\tau) \, = \, \frac{2}{3} \frac{|\tau|}{l} \, , 
\ee
we have
\begin{align}
P_{\mathcal{R}}(k, t)  & =  \frac{1}{\left( 2\pi l \right)^2}  k^3|\delta \phi _k \left( z,t \right)|^2 \nonumber \\
                         & = \frac{1}{\left( 2\pi l \right) ^2} k^3 \left| \int ds M(s,z) \mathcal{O} \left(k, t + s \right) \right|^2 \,.
\label{power_spectrum_bdry}
\end{align}
Thus, the power spectrum of ${\mathcal{R}}$ has the same slope as that of
the boundary operator ${\mathcal{O}}$. This
boundary operator given in this equation is known from the AdS/CFT correspondence:
The scalar field in the bulk corresponds to the expectation value of the trace of the
square of the field strength of a conformal field theory living on the boundary:
\be \label{corresp1}
{\mathcal{O}} \, = \, \langle \mathrm{tr} F^2 \rangle \, .
\ee
This is the same operator
whose evolution was studied in our previous paper \cite{Brandenberger:2016egn},
and in the following we will just briefly summarize the analysis which relates
the late time spectrum of ${\mathcal{O}}$ with the initial spectrum of $\delta \phi$.

In our case, the boundary conformal field theory is a $\mathcal{N}=4$ 
Super Yang Mills (SYM) theory in $3+1$-dimensions 
with a Yang-Mills coupling that varies in time, inherited from the time-dependent 
dilaton from the bulk. Given this theory, we can evaluate the operator, since 
\be
F_{\mu \nu} \, = \, 2 \partial_{[\mu}A_{\nu]} - i \left[A_\mu, A_\nu\right] \, .
\ee
We ignore the term with the commutator since this is subdominant in our analysis. We 
adopt Coulomb gauge ($\partial_i A^i = 0$) and set $A_0 = 0$.  Then the Fourier 
transform of the field strength tensor reduces to 
\be
\mathrm{Tr}[F^2 (k,t)]  \, = \, 2 \dot{A}_k^2-2(k_j A_{k_i} )^2 + 2 k_i A_{k_j} k^j A^{k_i} \, ,
\ee
where summation over the index $k$ is implied.
We are only interested in the infrared (IR) modes, where $k$ is small, so we can drop 
the last two term of the previous expression and thus obtain the approximate relation
\be \label{corresp2}
\mathrm{Tr}[F^2 (k,t)] \, = \, 2  \dot{A}_k^2 \, .
\ee
where once again summation over $k$ is implied.

The fundamental field of this theory is the vector field $A_\mu$, and this can be 
evolved in time, given its equation of motion in the boundary. So, the field 
$A_\mu (+t_b)$, necessary to calculate the operator $\mathcal{O} \left(x, t_b \right)$ 
can be evolved from a initial vector field through $t=0$, the time when the singularity 
happens. This time also corresponds to the place where the YM coupling vanishes 
and the theory becomes free.

Since the gauge theory becomes free, it could have been expected that the
gauge field fluctuations pass through $t = 0$ without any problem. However,
as discussed in \cite{Brandenberger:2016egn} this is not the case. In
terms of the original Fourier space modes $A(k, t)$ there is a branch cut
in the evolution equations, and in terms of the canonically normalized field
corresponding to $A(k, t)$ there is in fact a divergence. This divergence
corresponds to the blowup of particle production which is expected from
the point of view of the bulk theory, where the fluctuation modes obtain
infinite squeezing at $t = 0$. Hence, it is not surprising that at the level
of fluctuations the boundary theory at this point also becomes sick and 
infinite particle production occurs. This does not allow us to evolve the field 
passed $t = 0$. In order to be able perform this evolution, we imposed a 
regularization of the YM coupling, making it constant during a period 
$\left[-\xi, \xi \right]$, where $\xi$ is smaller than $t_b$, and matching the 
solutions (and their first derivatives) in the periods $t<-\xi$, $-\xi < t < \xi$ and 
$ t > \xi$.

We perform this matching in the boundary theory since in the bulk, 
at times $\xi = \sqrt{\alpha '}$, where $\alpha '$ is the string scale, the 
Ricci scalar reaches the string scale and the bulk supergravity description 
breaks down.  So, matching in the boundary can be performed at time $\xi$, 
much closer to the singularity, where the bulk theory is already in the strong 
limit. Since the theory in the boundary has no gravity, this matching is under 
much better control and goes closer to the singularity than what could be
done by working in the bulk.

With that, we can relate the solution of the field $A_\mu$ from early to 
late times past the singularity, first re-scaling the gauge field by 
$\tilde{A}_k (t)=e^{\phi /2} A_k (t)$ to obtain a canonically normalized
field. The analysis of \cite{Brandenberger:2016egn} yielded the result
\begin{align}
\tilde{A}_k (t)=|t|^{\frac{1}{2}} \left[ D_J ^{+} J_{\nu_g} \left( |kt| \right) + D_Y ^{+} Y_{\nu_g} \left( |kt| \right) \right]\,,
\end{align}
where the mode coefficients $D_J ^{+}$ and $D_Y ^{+}$  are related to the 
ones $D_J ^{-}$ and $D_Y ^{-}$before the singularity by:
\begin{align}
D_J ^{+} & = \frac{1}{2\nu_g} D_Y ^{-} + 2^{-2\nu_g} \left( 1+\frac{1}{2\nu_g} \right) \left( k\xi \right) ^{2\nu_g} D_J ^{-} \,, \\
D_Y ^{+} & =  - \frac{1}{2\nu_g} D_J ^{-} + 2^{2\nu_g} \left( 1- \frac{1}{2\nu_g} \right) \left( k\xi \right) ^{-2\nu_g} D_Y ^{-} \,.
\end{align}
At late times, at time $t=t_b$, when we map the results from the boundary to the 
bulk, the gauge field is then given by:
\begin{equation}
\tilde{A}_k(t_b) \simeq \left(  \frac{t_b}{\xi} \right)^{2 \nu_g} \tilde{A} \left( -t_b \right)\,,
\label{gauge_field_future}
\end{equation}
which means that the $k$-dependence if the field remains the same after 
passing through the singularity, changing only its amplitude that is enhanced
by the factor 
\be
\mathcal{F}(t) \, = \, \left(  t_b/\xi \right)^{2 \nu_g} \,  . 
\ee
So, given an initial condition in the gauge field $A_k (-t_b)$, we can time 
evolve this field until time $t_b$ in the future of the singularity, and then 
calculate the operator $\mathrm{Tr} \left[ F^2(k,t') \right]$ and obtain the power spectrum.

This initial value for the gauge field in the boundary theory can be inferred from the 
initial scalar field in the bulk. As done in our previous work, we choose a particular
scaling of the Fourier modes of the boundary gauge field such that the operator
$\mathcal{O}$ has the same amplitude and scaling as what is induced from the
bulk scalar field fluctuations which we are starting out with. From (\ref{corresp1}) and 
(\ref{corresp2}) we have
\begin{equation}
m_{pl}^4 \delta \phi^{in} (k) \, = \, 
\frac{1}{2} \int _0 ^k d^3 k' \dot{A}^{in} \left( \frac{k+k'}{2} \right) \dot{A}^{in}\left( \frac{k-k'}{2} \right) V^{1/2} \,
= \, \mathrm{Tr} \left[ F^2 (-t_b) \right]_k \, = \, \mathcal{O} \left( k, -t_b \right) \, ,
\label{matching_past}
\end{equation}
where $V$ is a normalization volume introduced in the definition of the Fourier transform (such
that the Fourier modes of $A$ have the mass dimension of a harmonic oscillator, {\it i.e.} $-1/2$). 
This integral can be performed in two regions, $R_1$ where $k < k'$, and $R_2$ where $k > k'$. 
In region $R_1$ we can set $k'=0$ and get, approximately
\be \label{contrib}
m_{pl}^4 \delta \phi^{in} (k) \, \sim \, k^3 \dot{A}_k^2 V^{1/2} \, .
\ee 

In the case when the initial bulk scalar field has a scale invariant power spectrum, 
$\delta \phi \propto k^{-3/2}$, the gauge field at $t=-t_b$, the time of matching 
onto the boundary in the past, has $A_k^{in} (t) = A_k (-t_b) \sim k^{-9/4}$. 
Assuming this scaling for the gauge Fourier modes, it can easily be seen that Region 
$R_2$ gives gives a contribution comparable to (\ref{contrib}). With the
initial conditions given by (\ref{matching_past}) and the growth of the gauge modes
given by (\ref{gauge_field_future}) we can write the boundary data in the future, 
encoded in the operator:
\begin{equation}
\mathcal{O} \left( k, t \right) \, = \, 
\frac{(4\nu_g)^2}{\xi^{4\nu_g}} t^{2(2\nu_g)} \frac{1}{2} \int _0 ^k d^3 k' \dot{A} \left( \frac{k+k'}{2}, -t_b \right) \dot{A}\left( \frac{k-k'}{2}, -t_b \right) V^{1/2} = \mathrm{Tr} \left[ F^2 (-t_b) \right]_k \, .
\end{equation}

Now we have all the ingredients to obtain the power spectrum of curvature perturbations 
past the singularity, given an initial bulk scalar field perturbation in the past:
\begin{align}
P_{\mathcal{R}}  &= \frac{k^3}{\left( 2\pi \right) ^2} \left| \int ds M(s,z) \mathrm{Tr} \left[ F^2 (t_b) \right]_k  \right|^2 \,, \nonumber \\
&=  \frac{\left( 4 \nu _g \right) ^4}{\left( 2\pi \right) ^2 \xi ^{8\nu_g}}  \, k^3 \,  t^{4(2\nu_g )}  \left| \int ds M(s,z) \frac{1}{2} \int _0 ^k d^3 k' \dot{A} \left( \frac{k+k'}{2}, -t_b \right) \dot{A}\left( \frac{k-k'}{2}, -t_b \right) V^{1/2} \right|^2 \,.
\label{PS_final}
\end{align}
Integrating over region $R_1$, and taking $k'=0$, we have
\begin{align}
P_{\mathcal{R}}  &= \frac{\left( 4 \nu _g \right) ^4}{\left( 2\pi \right) ^2 \xi ^{8\nu_g}}  \, k^3 \,  t^{4(2\nu_g )} \left| \int ds M(s,z) \, k^3 \dot{A}^2(k,-t_b) \right|^2 \,, \nonumber \\
&= \frac{\left( 4 \nu _g \right) ^4}{\left( 2\pi \right) ^2 \xi ^{8\nu_g}}  \, k^3 \,  t^{4(2\nu_g )} \left| \int ds M(s,z) m_{pl}^4 \delta \phi^{in} (-t_b) \right|^2 \,.
\label{PS_final}
\end{align}
For the case presented in \cite{Brandenberger:2016egn} when we have a scale invariant 
spectrum for the bulk scalar field in the past, with $\delta \phi^{in} \propto k^{-3/2}$, this 
implies that $\dot{A}(k, -t_b) \propto k^{-9/4}$. Plugging this expression 
into (\ref{PS_final}) we see that at $t=t_b$ in the future:
\begin{align}
P_{\mathcal{R}}  &\simeq 
\frac{\left( 4 \nu _g \right) ^4}{\left( 2\pi l \right) ^2 \xi ^{8\nu_g}} \,  t_b^{4(2\nu_g)} 
\bigl[ \int ds M(s,z) \bigr]^2 m_{pl}^{8} \, .
\end{align}
The power spectrum for the curvature perturbations is scale invariant. This means that 
the index of the power spectrum of the curvature flluctuations  is not 
changed after passing through the singularity. So, if we start in the contracting phase with a
scale invariant power spectrum of curvature fluctuations before the singularity, then the 
final curvature  perturbations will also be scale invariant, carrying at late times an 
enhancement factor in the amplitude, related to particle production occurring on super-Hubble
scales close to the bulk singularity. 

\section{Conclusions and Discussion}

We have used the AdS/CFT correspondence to propagate cosmological fluctuations from
the contracting phase to the expanding phase of a time-dependent deformation of an AdS
bulk space-time which has a curvature singularity at a time $t = 0$. The bulk space-time
is weakly coupled for $|t| > t_b$, and strongly coupled for $|t| < t_b$. Since the
CFT on the boundary becomes weakly coupled for $|t| < t_b$, we map the bulk 
perturbations onto the boundary at the transition time $t = - t_b$, evolve the fluctuations
in the conformal field theory until $t = t_b$, and then reconstruct the bulk perturbations.

We have shown that there is a gauge choice for the bulk space-time coordinates in
which the information about cosmological fluctuations can be encoded in terms of
the dilaton perturbations. This is the frame we use to map the inhomogeneities onto
the boundary. We use the same choice of coordinates to  reconstruct the bulk for
in the future of the strongly coupled bulk region, i.e. for $t > t_b$.

For the background with a bulk curvature singularity, particle production in the boundary
theory diverges at $t = 0$. Hence, to obtain a well-defined evolution, we need to regulate
the boundary theory (and thus also the bulk theory) in some time interval $|t| < \xi$,
where $\xi \ll t_b$. In the regulated theory, it is then possible to unambiguously compute
the evolution of the linearlized cosmological perturbations. We find that, as in the case of
dilaton fluctuations in \cite{Brandenberger:2016egn}, 
the spectral index of infrared perturbations is the same
before entering and after exiting the region of large space-time curvature. The
amplitude of the spectrum, on the other hand, changes by a factor which depends
on the ratio of $t_b$ and $\xi$. These results agree with what is obtained in some
models of nonsingular bounces in which rather ad hoc new physics is used to
obtain the bounce (see e.g. \cite{5dflucts}). What is satisfying in our approach is that
the bounce is obtained using fundamental ingredients from superstring theory.

There are other ways to obtain a bouncing cosmology from superstring theory.
One recent example makes use of the $T$-duality symmetry in the Euclidean
time direction to obtain a so-called S-brane bounce \cite{Sbrane}. Another
approach is in \cite{Edna}. It is also possible that as a consequence of the
Hagedorn spectrum of string states \cite{Hagedorn} coupled with the T-duality
symmetry in compact spatial directions one obtains an early emergent
Hagedorn phase \cite{BV}, in which case thermal fluctuations of a gas
of strings would be the source of the observed cosmological perturbations \cite{NBV}.
These different approaches lead to signatures which are distinguishable (and
also distinguishable from conventional inflationary cosmology) in cosmological
observations, in particular because of different predictions of the tilt of
the gravitational wave spectrum \cite{BNPV}, the running of the
spectrum \cite{Edward} and the amplitude and shape of the three point
function of the curvature fluctuation \cite{fnl}.

\begin{acknowledgments}

The authors would like to thank Yifu Cai, Sumit Das, Alberto Enciso, Niky Kamran, Yi Wang
and in particular Ian Morrison for fruitful discussions. E.F. acknowledges financial
support from CNPq (Science without Borders). RB wishes to thank the Institute for 
Theoretical Studies (ITS) of the ETH Z\"urich for kind hospitality. He acknowledges financial 
support at the ITS from Dr. Max R\"ossler, the Walter Haefner Foundation and the 
ETH Zurich Foundation. The research of RB is also supported in
part by funds from NSERC and the Canada Research Chair program, and by 
a Simons Foundation fellowship.

\end{acknowledgments}


\end{document}